\documentclass[aps,prd,twocolumn,floatfix,noshowpacs,tightenlines,noshowkeys,superscriptaddress,amsmath,amssymb,nofootinbib]{revtex4}
\usepackage{amssymb,amsbsy,epsfig,color,graphicx}
\usepackage{color}
\usepackage{[longtable}
\usepackage{array}
\usepackage{dcolumn}   
\usepackage{cellspace}
\usepackage{mathtools}
\usepackage{amstext}
\usepackage{amssymb}
\usepackage{stmaryrd}
\usepackage{stackrel}
\usepackage{graphicx}
\usepackage{esint}
\usepackage[utf8]{inputenc}
\usepackage{blindtext}
\usepackage{float}
\restylefloat{table}
\usepackage{booktabs}
\usepackage{enumitem} 

\usepackage{etoolbox} 
\usepackage{lipsum} 
\usepackage[capitalize]{cleveref}
\usepackage{multirow}
\usepackage[caption=false]{subfig}
\renewcommand\[{\begin{equation}}
\renewcommand\]{\end{equation}}

\newcommand{\ba}{\begin{eqnarray}}
\newcommand{\ea}{\end{eqnarray}}

\makeatletter

\appto{\appendix}{%
\@ifstar{\def\theequation@prefix{A.}}%
{}%
}
\makeatother

\begin{document}

\title{Nonlocal star as a blackhole mimicker}

\author{Luca Buoninfante}
\affiliation{Dipartimento di Fisica "E.R. Caianiello", Universit\`a di Salerno, I-84084 Fisciano (SA), Italy}
\affiliation{INFN - Sezione di Napoli, Gruppo collegato di Salerno, I-84084 Fisciano (SA), Italy}
\affiliation{Van Swinderen Institute, University of Groningen, 9747 AG, Groningen, The Netherlands}


\author{Anupam Mazumdar}
\affiliation{Van Swinderen Institute, University of Groningen, 9747 AG, Groningen, The Netherlands}


\begin{abstract}
In the context of ghost-free, {\it infinite derivative gravity}, we will provide a quantum mechanical framework in which we can describe astrophysical objects devoid of curvature singularity and event horizon. In order to avoid ghosts and singularity, the gravitational interaction has to be nonlocal, therefore, we call these objects as {\it nonlocal stars}. Quantum mechanically a nonlocal star is a self-gravitational bound system of many gravitons interacting nonlocally. Outside the nonlocal star the spacetime is well described by the Schwarzschild metric, while inside we have a non-vacuum spacetime metric which tends to be conformally flat at the origin. Remarkably, in the most compact scenario the radius of a nonlocal star is of the same order of the Buchdahl limit, therefore slightly larger than the Schwarzschild radius, such that there can exist a photosphere. These objects live longer than a Schwarzschild blackhole and they are very good absorbers, due to the fact that the number of available states is larger than that of a blackhole. As a result nonlocal stars, not only can be excellent blackhole mimickers, but can also be considered as dark matter candidates. In particular, nonlocal stars with masses below $10^{14}$g can be made stable compared to the age of the Universe.
\end{abstract}

\maketitle


\section{Introduction}\label{intro}

Einstein's general relativity (GR) has given the best mathematical description of the gravitational interaction  and its predictions have been tested to a very high precision in the infrared (IR) regime, i.e. at large distances and late times \cite{-C.-M.,-B.-P.}. Despite these great achievements, there are still open questions regarding short distances and small time scales, i.e. the ultraviolet (UV) regime, where the gravitational interaction is least known. From an experimental point of view, Newton's inverse square law has been tested up to roughly $5$ micrometers \cite{-D.-J.}, which translates to roughly  $0.001$eV. Beyond these energies, the gravitational interaction has been hardly constrained so far with direct experiments.

From a conceptual point of view, Einstein's GR suffers from the presence of classical (blackhole and cosmological) singularities~\cite{Hawking}, and at a quantum level it turns out to be non-renormalizable beyond one loop, thus lacking predictability at high energies~\cite{tHooft:1974toh}. Moreover, one of the most important feature of blackholes in GR is the presence of event horizons, which has lead to many confusions and paradoxes, i.e. Hawking's information loss paradox~\cite{Hawking:1974sw}.

An interesting observation was made in Ref.\cite{-K.-S.} that by extending the Einstein-Hilbert action with the addition of quadratic curvature invariants, the renormalizability issue can be solved, but it was also shown that such a theory possesses a spin-$2$ massive ghost, which causes Hamiltonian instabilities at the classical level and unitarity violation at the quantum level. Furthermore, such finite derivative theories still harbor cosmological and blackhole singularities.

Recently, it  has been pointed out  that a possible way to ameliorate the issue of ghost is to go beyond finite order derivative theories, and to modify the action by introducing differential operators made up of infinite order covariant derivatives. Such an action indeed gives rise to nonlocal graviton interactions. Already in the early fifties,  it was realized that the UV behavior of loop integrals in quantum field theory could be ameliorated by working with nonlocal actions \cite{Yukawa:1950eq}, in particular generalizing local theories by introducing exponentials of an entire functions in the action. These nonlocal models were also studied from an axiomatic point of view, see Ref.\cite{efimov}.

First applications of this kind of nonlocal operators in the context of gauge theories and gravity were considered in Refs.\cite{Krasnikov,Kuzmin}, respectively. Subsequently, such theories were rediscovered in Ref.\cite{Tomboulis:1997gg,Moffat} where it was shown that higher derivative gauge theories can be made ghost-free.  In the gravitational context the same class of nonlocal actions were useful to obtain ghost-free nonsingular cosmological \cite{Biswas:2005qr,Biswas:2010zk,Biswas:2012bp,Koshelev:2012qn,Koshelev:2018rau} and blackhole \cite{Biswas:2011ar,Modesto:2011kw,Biswas:2013cha,Edholm:2016hbt,Frolov:2015bia,Frolov,Frolov:2015usa,Koshelev:2017bxd,Buoninfante:2018xiw,Koshelev:2018hpt,Buoninfante:2018rlq,Buoninfante:2018stt,Buoninfante:2018xif,Boos,Kilicarslan:2018yxd} solutions. Such nonlocal models were also applied in the framework of inflationary cosmology \cite{inflation}, and thermal field theory \cite{Biswas:2009nx,Biswas:2010xq,Biswas:2010yx}.

Field theoretical studies aimed to understand the UV behavior and to prove perturbative unitarity of ghost-free nonlocal field theories were made in Refs.\cite{Modesto:2011kw,okada,Tomboulis:2015gfa,Gama:2018cda,Hashi:2018kag,Talaganis:2014ida,Ghoshal:2017egr,Buoninfante:2018mre,Buoninfante:2018lnh} and in Refs.\cite{sen-epsilon,carone,chin,Briscese:2018oyx}, respectively. Very interestingly, the same kind of form factors also appear in the context of string field theory \cite{Witten:1985cc,eliezer,Tseytlin:1995uq,Siegel:2003vt}, where the cubic vertex of the open string tachyon field is nonlocal and given by $(e^{\Box/M_s^2}\phi)^3,$  and in $p$-adic string \cite{Freund:1987kt} where the kinetic operator is given by $e^{\Box/M_s^2}$, with $\Box$ being the d'Alembertian operator and $M_s$ denoting the string scale or, in other words, the scale of nonlocality.

The aim of this paper is to provide a quantum description of a massive, non-singular, horizon-less compact astrophysical object in the context of a ghost free, infinite derivative theory of gravity. In Ref.~\cite{Koshelev:2017bxd} it was already pointed out that an interesting possibility arises, where in order to preserve the {\it Area-Law} of gravitational entropy, the scale of nonlocality may transmute towards the IR regime. This has recently been checked at a quantum level by studying the scattering diagram of a system of $N$ gravitons interacting nonlocally, and forming a bound system very similar to a Bose-Einstein condensate~\cite{Buoninfante:2018gce}. In this paper, for the first time we will provide the spacetime metric of such a self gravitating system, which we call here as a {\it nonlocal star}. For such an object, we will compute the life time, the entropy and the number of Bekenstein states, and study the phenomenological consequences for astrophysics and cosmology.

The paper is organized as follows. In Section \ref{class-asp}, we will provide a quantum mechanical  treatment of nonlocal gravitational interaction.  In particular, we will study the properties of a condensate made up of $N$ (scalar) gravitons and show the existence of a {\it complementarity relation} between the total mass of a nonlocal condensate and the effective scale of nonlocality, which is the key for the absence of a horizon. In Section \ref{sec-condens}, we will make a qualitative estimation of important classical and quantum physical quantities such as compactness, life time, entropy, absorption and reflection coefficients of a nonlocal star. In Section \ref{concl}, we will discuss the outlook and draw the conclusions. 


\section{Quantum aspects of nonlocal gravitational interaction}\label{class-asp}

\subsection{Infinite derivative gravitational action}

The most general quadratic action in $4$ dimensions, which is parity invariant and torsion-free is given by \cite{Biswas:2011ar,Biswas:2016etb}\footnote{It is worthwhile to mention that the most general quadratic gravitational action which contains torsion and generalizes the Poincar\'e gravity has been constructed recently in Ref.\cite{delaCruz-Dombriz:2018aal}. In \cite{Biswas:2011ar,Biswas:2016etb} the quadratic curvature action was constructed systematically based on diffeomorphism invariance and preserving the ghost free conditions. See also Ref.\cite{Mazumdar:2018xjz} for a $3$D version of the gravitational action in Eq.\eqref{quad-action}.}:
\begin{equation}
\begin{array}{rl}
S=& \displaystyle \frac{1}{16\pi G}\int d^4x\sqrt{-g}\left\lbrace \mathcal{R}+\beta\left(\mathcal{R}\mathcal{F}_1(\Box_s)\mathcal{R}\right.\right.\\[3mm]
& \displaystyle  \,\,\,\,\,\,\,\,\,\,\,\,\,\,\,\,\left.\left.+\mathcal{R}_{\mu\nu}\mathcal{F}_2(\Box_s)\mathcal{R}^{\mu\nu}+\mathcal{R}_{\mu\nu\rho\sigma}\mathcal{F}_3(\Box_s)\mathcal{R}^{\mu\nu\rho\sigma}\right)\right\rbrace,
\end{array}
\label{quad-action}
\end{equation}
where $\mu,~\nu=0,1,2,3$; we will work mostly with positive signature $(-+++).$ Moreover, we will use the physical units in which $c=1,$ but $\hbar\neq 1,$ so that classical and quantum aspects can be explicitly distinguished, thus $G=\hbar/M_p^2=L_p^2/\hbar$ is the Newton constant and $\beta= L_s^2/2=\hbar^2/(2M_s^2)$ is a dimensionful coupling, with $M_s$ being the fundamental scale of nonlocality, which in the context of string theory corresponds to the string scale. The three gravitational form-factors $\mathcal{F}_{i}(\Box_s)$ are covariant functions of the d'Alembertian and can be uniquely determined around the Minkowski background~\cite{Biswas:2011ar,Biswas:2013cha}. By setting $\mathcal{F}_3(\Box_s)=0,$ for simplicity and without any loss of generality up to quadratic order in the metric perturbation around flat background, we can keep the massless spin-$2$ graviton as the only dynamical degree of freedom by imposing the following condition~\footnote{In this paper we will only consider analytic form-factors. However, it is worth mentioning that nonlocal models with non-analytic differential operators have been investigated by many authors; see, for example, Refs. \cite{Bravinsky,Deser:2007jk,Conroy:2014eja,Belgacem:2017cqo,Woodard:2018gfj,belenchia}.}:
$2\mathcal{F}_1(\Box_s)=-\mathcal{F}_2(\Box_s)$ as shown in Ref.\cite{Biswas:2011ar} around the Minkowski background.
By expanding around Minkowski, $$g_{\mu\nu}=\eta_{\mu\nu}+\kappa h_{\mu\nu},$$ with $\kappa:=\sqrt{8\pi G},$ we obtain
\begin{equation}
S=\frac{1}{4}\int d^4x\, h_{\mu\nu}\,(1-\mathcal{F}_1(\Box_s)\Box_s)\mathcal{O}^{\mu\nu\rho\sigma}\,h_{\rho\sigma}+\mathcal{O}(\kappa h^3),\label{lin-action}
\end{equation}
where $\mathcal{O}(\kappa h^3)$ takes into account of higher order terms in the perturbation, while the four-rank operator $\mathcal{O}^{\mu\nu\rho\sigma}$ is totally symmetric in all its indices and defined as
\begin{equation}
\begin{array}{ll}
\displaystyle \mathcal{O}^{\mu\nu\rho\sigma}:= \displaystyle \frac{1}{4}\left(\eta^{\mu\rho}\eta^{\nu\sigma}+\eta^{\mu\sigma}\eta^{\nu\rho}\right)\Box-\frac{1}{2}\eta^{\mu\nu}\eta^{\rho\sigma}\Box&\\
\,\,\,\,\,\,\,\,\,\,\,\,\,\,\,\,\,\displaystyle +\frac{1}{2}\left(\eta^{\mu\nu}\partial^{\rho}\partial^{\sigma}+\eta^{\rho\sigma}\partial^{\mu}\partial^{\nu}-\eta^{\mu\rho}\partial^{\nu}\partial^{\sigma}-\eta^{\mu\sigma}\partial^{\nu}\partial^{\rho}\right).&
\end{array}\label{4-rank-oper}
\end{equation}
By inverting the kinetic operator we obtain the graviton propagator around the Minkowski background, and its saturated and gauge independent part is given by \cite{Biswas:2011ar,Biswas:2013kla}
\begin{equation}
\Pi_{\mu\nu\rho\sigma}(k)=\frac{1}{1+\mathcal{F}_1(k)k^2/M_s^2}\left(\frac{\mathcal{P}_{\mu\nu\rho\sigma}^2}{k^2}-\frac{\mathcal{P}^0_{s,\,\mu\nu\rho\sigma}}{2k^2}\right),\label{propag}
\end{equation}
where $\mathcal{P}^2/k^2-\mathcal{P}^0_{s}/2k^2$ is the graviton propagator of Einstein's GR, while $\mathcal{P}^2$ and $\mathcal{P}^0_s$ are two spin projection operators projecting along the spin-$2$ and spin-$0$ components, respectively; see Refs.\cite{Biswas:2013kla,Buoninfante} for further details. Note that in order not to introduce any extra dynamical degrees of freedom other than the massless spin-$2$ graviton, we need to require that the function $1+\mathcal{F}_1(k)k^2/M_s^2$ does not have any zeros, i.e. that it is an {\it exponential of an entire function}:
\begin{equation}
1+\mathcal{F}_1(k)\frac{k^2}{M_s^2}=e^{\gamma(k^2/M_s^2)},\label{choice}
\end{equation}
where the $\gamma(k^2/M_s^2)$ is an entire function. We will mainly work with the simplest choice $\gamma(k^2)=k^2/M_s^2$, but we will also discuss other kind of entire functions as well in Appendix \ref{app-transmut}; see also Ref.~\cite{Edholm:2016hbt} for other examples of entire functions.


\subsection{Cubic graviton interaction in infinite derivative gravity}

We wish to study the gravitational interaction, up to cubic order in the metric perturbation $h_{\mu\nu}.$ Due to the complicated structure of the graviton vertices in the infinite derivative gravity, for simplicity we will work {\it only} with the scalar component of the graviton, i.e. with the trace field $h,$ and consider interaction vertices up to $\mathcal{O}(\kappa h^3)$. All possible interaction terms which can be constructed are the following \cite{Talaganis:2014ida}:
\begin{equation}
\begin{array}{rl}
S_h^{(3)}\sim &\displaystyle \kappa \int d^4x\left(a\,h\partial_{\mu}h\partial^{\mu}h+b\,h\partial_{\mu}he^{-\Box/M_s^2}\partial^{\mu}h\right.\\
&\displaystyle \,\,\,\,\,\,\,\,\,\,\,\,\,\,\,\,\,\,\,\,\,\,\,\left.+c\,h\Box he^{-\Box/M_s^2}h\right), \label{cubic-graviton interac}
\end{array}
\end{equation}
where $a,b,c$ are three constant parameters which need to be fixed. In Ref.\cite{Talaganis:2014ida}, they were fixed to $a=b=-c=1/4$ by demanding the action around the Minkowski vacuum to be invariant under the infinitesimal scaling transformation: $g_{\mu\nu}\longrightarrow (1+\varepsilon) g_{\mu\nu}$ that translates into $h \longrightarrow (1+\varepsilon)h+\varepsilon.$ 

We would like to emphasize that in this manuscript we will consider the gravitons composing a condensate confined in a bound state by their own self-gravity, this means that such gravitons are not the ones with $+2$ and $-2$ helicity which propagate on-shell, but they are off-shell. An off-shell graviton has six degrees of freedom coming from a spin-$2$ (five components) and a spin-$0$ (one component); see for instance Eq.\eqref{propag}. Therefore, the use of only the scalar component of the propagator, as a simpler case, is still good enough to understand which are the main implications due to nonlocality.

\subsection{$N$ gravitons interacting nonlocally} \label{quant-asp}

Classically, $M_s$ (or $L_s$) is a fundamental parameter, but quantum mechanically this becomes a dynamical quantity as it has been recently shown in Ref.\cite{Buoninfante:2018gce}. It was realized that for a system of $N$ scalar gravitons interacting nonlocally, the nonlocal energy scale can transmute to lower energies.  For instance, for an $N$-point scattering amplitude in the limit $N\gg 1,$ it was shown the following behaviour \cite{Buoninfante:2018gce}~\footnote{See Appendix \ref{app-transmut} for more details.}:
\begin{equation}
\mathcal{M}_N\sim 
\frac{e^{-N^{3}k^2/M_s^2}}{k^{2N}}= \frac{e^{-K^2/M_{\rm eff}^2}}{k^{2N}}, \label{N-point-zero-ext}
\end{equation}
where we have neglected constant factors for simplicity. From Eq.\eqref{N-point-zero-ext}, by defining $K=Nk$ to be the total momentum (energy), it is clear that the $N$-point amplitude depends on the following {\it effective} nonlocal scale~\cite{Buoninfante:2018gce}:
\begin{equation}
M_{\rm eff}= \frac{M_s}{\sqrt{N}},\,\,\, {\rm or}\,\,\,L_{\rm eff}= \hbar M_{\rm eff}^{-1}=\sqrt{N}L_s, \label{effective-energy}
\end{equation}
meaning that the scale at which the nonlocal effects become relevant is {\it not} fixed, but dynamical, and depends on the number $N$ of interacting scalar gravitons. In terms of length scales, this means that the higher is the number of interacting quanta, the larger will be the nonlocal region on which the interaction happens. 
Note that a similar scaling behavior was also obtained in Ref.\cite{Koshelev:2017bxd} from a different point of view by demanding that the gravitational entropy of a self-gravitating system preserves the  {\it Area-law}~\cite{Koshelev:2017bxd}. 

The behavior in Eq.\eqref{N-point-zero-ext} refers to amplitudes with zero-external momenta, which correspond to bound systems whose constituents can be seen as weakly interacting off-shell quanta. In fact, a Bose-Einstein condensate is a system of weakly coupled bosons whose number can be very large, but still allowing all the constituents to live in the ground-state. There is no external interaction (external legs) which can excite the condensate in such a way that quanta can escape from the ground-state, but only internal interactions among the constituents. Moreover, it is well known that a condensate exhibits a collective behavior where all the constituents have a wave-length of the order of the size of the system, which in our case is $L_{\rm eff}=\hbar M_{\rm eff}^{-1}$. 


\section{Nonlocal star as a condensate of gravitons}\label{sec-condens}

In this Section we wish to study some of the quantum properties of a Bose-Einstein condensate made up of gravitons interacting nonlocally. Note that condensates of attractive bosons and their phase transitions were first studied in Ref.\cite{Bogolyubov:1947zz}, while a pioneering application to the case of gravitons and blackhole physics was worked out in Ref.\cite{Dvali:2011aa}. In this paper we will generalize their treatment to the case of a system of $N$ gravitons interacting nonlocally, which physically represent a new astrophysical object called {\it nonlocal star}.

Let us assume that each quanta (graviton) bring an energy (or effective mass), $E_g,$ and so a wave-length
\begin{equation}
\lambda_{g}=\frac{\hbar}{E_g}. \label{wave-length-single}
\end{equation}
Each individual graviton feels nonlocal interaction if and only $\lambda_g\sim L_s,$ while for wave-lengths $\lambda_g>L_s$ the self-graviton interaction is just local. Therefore, a nonlocal graviton is characterized by a wave-length of the order of the fundamental scale of nonlocality:
\begin{equation}
\lambda_g\sim L_s\sim \hbar M_s^{-1},\,\,\,\,\,\,\,\,\,{\rm or\,\,\, equivalently,}\,\,\,\,\,\,\,\,\,E_g\sim M_s\label{wave-length_mass-single}\,.
\end{equation}
Given a weakly interacting system, we can define a dimensionless quantum self-coupling for gravitons as follows:
\begin{equation}
\alpha_g=\frac{\hbar G}{\lambda_g^2}=\frac{L_p^2}{L_s^2}=\frac{M_s^2}{M_p^2}<1, \label{self-coup-less-1}
\end{equation}
where the inequality always holds true, since  $M_s < M_p$ (or $ L_s>L_p).$

For a gravitational system of $N$ nonlocal gravitons, the interaction happens on a region of size $\lambda \sim R\sim L_{\rm eff}= \hbar M_{\rm eff}^{-1}$ and a collective quantum coupling can be defined as follows~\footnote{A local version of this graviton condensate was applied to blackholes by the authors in Refs.\cite{Dvali:2011aa}, and further studied in many other works; see for example Refs.\cite{Dvali:2012en,Dvali:2013vxa,Dvali:2013eja,Dvali:2014ila,Dvali:2015aja,Casadio:2014vja,Casadio:2015bna,Casadio:2016zpl,Casadio:2016aum,Buoninfante:2019fwr}. In this case, a blackhole is assumed to be a leaky Bose-Einstein condensate of attractive off-shell longitudinal gravitons stuck at the critical point of a phase transition, and all the physical properties can be uniquely determined once the number $N$ of gravitons is given. In the quantum blackhole picture we have $R\sim r_{\rm sch}$, therefore for $N$ gravitons we obtain $\lambda_g^2 \sim NL_p^2,$ in such a way that the collective quantum coupling is always equal to one, $N\alpha_g=1,$ which means that the gravitational system is stuck at its critical point \cite{Dvali:2012en}.}:
\begin{equation}
N\alpha_g=N\frac{L_p^2}{L_{\rm eff}^2}=\frac{L_p^2}{L_s^2}=\frac{M_s^2}{M_p^2}<1. \label{self-coup-less-nonlocal}
\end{equation}
which is still less than one. From the inequalities in Eqs.(\ref{self-coup-less-1},\ref{self-coup-less-nonlocal}) it is evident that at the quantum level nonlocality weakens the strength of the gravitational interaction as the quantum coupling turns out to be less than one. Very interestingly, by making a comparison with the quantum blackhole picture in Einstein's GR, see Ref.\cite{Dvali:2011aa}, a nonlocal star never reaches the critical point $N\alpha_g=1.$ This feature is a crucial point in order to avoid the formation of any horizon, as we will discuss below.


\subsection{Mass of a nonlocal star}\label{condens-subsec}

Given $N$ nonlocal gravitons of energy $E_g\sim M_s,$ the total mass of a gravitational system will be given by: 
\begin{equation}
m =NE_g\,. \label{total mass.}
\end{equation}
We now know that the wave-length of each nonlocal graviton becomes larger due to the collective behavior of the condensate, as described in Eq.\eqref{effective-energy}, so that the following scaling behavior for each quanta will hold:
\begin{equation}
\lambda_g\sim \sqrt{N}L_s\sim L_{\rm eff}.\label{wave-length2}
\end{equation}
In terms of the mass/energy, each  graviton becomes {\it softer}
\begin{equation}
E_g\sim \frac{M_s}{\sqrt{N}}\label{soft-mass}.
\end{equation}
Note that the quantity in Eq.\eqref{wave-length2} corresponds to the size of the system, which is approximatively given by $L_{\rm eff}=\sqrt{N}L_s.$ From Eq.\eqref{soft-mass} we can now understand that the total mass of the graviton condensate in Eq.\eqref{total mass.} reads
\begin{equation}
m=NE_g \sim N \frac{M_s}{\sqrt{N}}=\sqrt{N}M_s\label{total mass.2}.
\end{equation}
%


\subsection{Gravitational potential of a nonlocal star}\label{lin-sol-sec}

We now wish to understand what is the gravitational potential inside and outside a nonlocal star.  
We have learned that all the constituents of the condensate feel a nonlocal interaction at the effective scale $M_{\rm eff}$ (or $L_{\rm eff}$). Therefore, by working in a static regime, $k^2\simeq \vec{k}^2,$ we can compute the gravitational potential felt at a point inside the system, due to each individual gravitons described by the stress-energy tensor $T_1^{\mu\nu}\sim E_g \delta^{\mu}_0\delta^{\nu}_0\delta^{(3)}(\vec{r}'_i)$, with $i=1,\dots,N\,.$ We can compute it by using the tree-level scattering amplitude technique, between $N$ sources $T1$ and the source of unit mass $T_2^{\mu\nu}\sim  \delta^{\mu}_0\delta^{\nu}_0\delta^{(3)}(\vec{r}),$ where $r'$ and $r$ are constrained to be inside the region of nonlocality, i.e. 
$r\sim r_{NL}$:
\begin{equation}
\!\!\!\begin{array}{rl}
\Phi(r)=& \displaystyle \sum\limits_{i=1}^N \Phi(\vec{r}-\vec{r}'_i)\\
\approx &\displaystyle -\kappa^2 \sum\limits_{i=1}^N \int\frac{d^3|\vec{k}|}{(2\pi)^3}T_1^{00}(k)\Pi_{0000}(k)T_2^{00}(-k)e^{i\vec{k}\cdot(\vec{r}-\vec{r}'_i)}\\
= & \displaystyle -\frac{\kappa^2 E_g}{2} \sum\limits_{i=1}^N\int\frac{d^3|\vec{k}|}{(2\pi)^3}\frac{e^{-\vec{k}^2/M_{\rm eff}^2}}{\vec{k}^2}e^{i\vec{k}\cdot(\vec{r}-\vec{r}'_i)},
\end{array}\label{exch}
\end{equation}
which, by assuming that each graviton contribute equally to the total gravitational potential at $r,$ becomes
\begin{equation}
\begin{array}{rl}
\Phi(r)\approx & \displaystyle -\frac{\kappa^2 NE_g}{2} \int\frac{d^3|\vec{k}|}{(2\pi)^3}\frac{e^{-\vec{k}^2/M_{\rm eff}^2}}{\vec{k}^2}e^{i\vec{k}\cdot \vec{r}}\\
=& \displaystyle -\frac{G m}{r}{\rm Erf}\left(\frac{M_{\rm eff}r}{2\hbar}\right)= \displaystyle-\frac{G m}{r}{\rm Erf}\left(\frac{r}{2 L_{\rm eff}}\right).
\end{array}\label{potential-graviton-exch}
\end{equation}
%
%
%
Outside the nonlocal star, we have a vacuum, and the spacetime is well described by the Schwarzschild metric, where $\Phi\sim -Gm/r.$ In fact, outside the nonlocal region the scale of nonlocality is given by $M_s=\hbar L_s^{-1},$ and for any point $r\gg L_s,$ we have practically GR as an 
excellent approximation; see also Ref. \cite{Edholm:2016hbt}. 

\subsection{Radius of a nonlocal star}

From the result in Eq.\eqref{potential-graviton-exch}, we can also understand that the linear regime holds true as long as the gravitational metric potential satisfies the inequality $\kappa h_{00}=-2\Phi <1$ for any $r,$ which also means $2Gm/(\sqrt{\pi}L_{\rm eff})<1.$ By defining the radius of the nonlocal star as $r_{\rm NL}=2L_{\rm eff},$ the previous inequality can be also recast as follows:
\begin{equation}
r_{\rm NL}\sim 2L_{\rm eff}=r_{\rm sch}(1+\epsilon)\gtrsim 2 Gm\frac{2}{\sqrt{\pi}}. \label{inequality}
\end{equation}
where $r_{\rm sch}=2Gm$, and 
\begin{equation}
\epsilon \gtrsim  0.128\,,\label{epsilon}
\end{equation}
which is saturated in the most compact scenario, i.e. $\epsilon \simeq 0.128.$ From the inequality in Eq.\eqref{inequality}, we can understand that the radius of a nonlocal star always engulfs the Schwarzschild radius, which implies that there is no horizon. In this respect there will be no Hawking entangled pair production near $2Gm$, since the spacetime near $r_{\rm sch}$ is not sufficiently stretched~\cite{Hawking:1974sw}, and the light cone structure does not alter anywhere in the spacetime region. A Similar situation arises in the fuzz-ball scenario, which has been constructed in a stringy scenario, for a review see Ref.~\cite{Mathur:2005zp,Mathur:2017fnw}.

In fact, lack of horizon and singularity is true for any range of mass for the nonlocal star, provided the inequality in Eq.(\ref{inequality}) is satisfied. By knowing the expressions of the total mass and the effective nonlocal scale, see Eq.\eqref{total mass.2} and Eq.\eqref{effective-energy}, we find that the $2|\Phi| <1$, always holds true. Indeed, we have:
\begin{equation}
m M_{\rm eff}\sim \sqrt{N}M_s \frac{M_s}{\sqrt{N}}=M_s^2<M_p^2. \label{inequality-always}
\end{equation}
Increasing the mass of a gravitational system also means increasing the number of interacting gravitons, which in turn means shifting the nonlocal scale towards the infrared regime, in such a way that the inequality in Eq.\eqref{inequality} is always satisfied for any value of the mass.  This should be seen as a {\it complementary principle} for nonlocal systems~\footnote{It is worthwhile to mention that in order to avoid the horizon, a necessary condition is given by the relation in Eq.\eqref{self-coup-less-nonlocal} for the collective quantum coupling. Indeed, the blackhole picture in GR, given by Ref.\cite{Dvali:2011aa}, yields $N\alpha_g=1$ and, in our case, this  can be recovered when $M_s=M_p,$ which implies that the inequality in Eq.\eqref{inequality} can not be satisfied anymore, meaning that $\kappa h_{00}=1,$ i.e. gravity becomes strong enough to form a trapped surface, and eventually an event horizon.}.

Furthermore, note that the inequality in Eq.\eqref{inequality} is saturated for the most compact nonlocal star. However, depending on $M_s$ and $N$, the radius of the nonlocal star can be made larger than this value. In the latter case, the nonlocal star can swell up even beyond $3Gm$, where the photosphere of a Schwarzschild blackhole is. We will discuss this situation below.

\subsection{Metric of a nonlocal star }

The spacetime metric for the nonlocal star can be modelled as follows. We have seen that the inside of the star is a non-vacuum region in which nonlocal interactions among the constituents take place, in such a way that the overall metric potential $2|\phi|$ remains bounded by one. While, outside we have a vacuum region whose spacetime geometry is well described by the Schwarzschild metric. Therefore, we can construct the metric for a nonlocal star as follows~\cite{Koshelev:2017bxd,Buoninfante:2018rlq},
\begin{equation}
ds^2=-(1+2\Phi)dt^2+\frac{dr^2}{1+2\Psi(r)}+r^2d\Omega^2, \label{metric-non-loc-star}
\end{equation}
where\footnote{Note that the metric potentials in Eqs.(\ref{phi-pot},\ref{psi-pot}) are discontinuous at $r=r_{\rm NL}.$ However, there is no sharp boundary between nonlocal and local regimes, but the metric has to be smooth all the way from $r=\infty$ to $r=0;$ therefore, a more rigorous description should be able to take into account this issue in such a way that the there exist two metric potentials which continuously interpolate between the inner and outer regions. One way to model the above metric as a continuous function of the radial coordinate is to introduce two normalization constants in the denominators of the two metric potentials. Indeed, by writing $\Phi\rightarrow \Phi/a$ and $\Psi\rightarrow \Psi/b$ for $r\lesssim r_{\rm NL},$ where $a\equiv {\rm Erf}\left(1\right)$ and $b\equiv {\rm Erf}\left(1\right)-1/(\sqrt{\pi}e),$ the metric can be made continuous at $r=r_{\rm NL}.$ By modelling the metric in this fashion might be helpful for numerical computations.}
\begin{equation}
\displaystyle \Phi(r)=\left\lbrace \begin{array}{ll}
\displaystyle -\frac{G m}{r}{\rm Erf}\left(\frac{r}{r_{\rm NL}}\right), & \displaystyle r\lesssim r_{\rm NL}, 	\\
\displaystyle -\frac{Gm}{r}, & \displaystyle r>r_{\rm NL}, 	
\end{array}\right. \label{phi-pot}
\end{equation}
and
\begin{equation}
\!\!\displaystyle \Psi(r)=\left\lbrace \begin{array}{ll}
\displaystyle -\frac{G m}{r}{\rm Erf}\left(\frac{r}{r_{\rm NL}}\right)+\frac{2Gm\,e^{-r^2/r_{\rm NL}^2}}{\sqrt{\pi} r_{\rm NL}}, & \displaystyle r\lesssim r_{\rm NL}, 	\\
\displaystyle -\frac{Gm}{r}, & \displaystyle r>r_{\rm NL}. 	
\end{array}\right. \label{psi-pot}
\end{equation}
For such a metric all curvature invariants are non-singular, and it approaches conformal-flatness in the limit $r\longrightarrow 0$~\cite{Buoninfante:2018xiw,Buoninfante:2018rlq}:
\begin{equation}
ds^2\approx -d\tau^2+dr^2+r^2d\Omega^2,~~\tau=\sqrt{1+2A}\,t\,,
\end{equation}
with $2A=4Gm/(\sqrt{\pi}r_{\rm NL})< 1$. Moreover, for the inside metric the Birkhoff thereom is violeted, $\Phi\neq \Psi.$ Note that, the metric potential $\Phi$ is non-vanishing at $r=0$ meaning a net distinction between the nonlocal star and gravastar \cite{Mazur:2001fv,Mazur:2004fk} metrics, indeed the latter has a de Sitter core.

The nonlocal star is very similar to a blackhole in terms of an equation of state parameter. The object is held purely due to the energy density, the pressure component is zero in our case, i.e. $p=0$, which is very different from a boson star \cite{Schunck:2003kk}, or a neutron star \cite{Ozel:2016oaf}. However, there might be an effective pressure not related to the matter sector but coming from the modification to the spacetime geometry induced by nonlocality. There is some resemblance of these objects with non-commutative geometry, see~\cite{Nicolini:2005vd,Nicolini:2008aj}, in terms of resolving the blackhole singularity problem.


\subsection{Nonlocal star with or without a photosphere}\label{compact-section}

Let us define the following function
\begin{equation}
\mu:=1-\frac{r_{\rm sch}}{r_{\rm NL}}=\frac{\epsilon}{1+\epsilon},
\label{function-compact}
\end{equation}
which measures the compactness of a nonlocal star.
For a blackhole we have $\epsilon=0,$ which implies $\mu_{\rm bh}=0.$ While, in the case of the most compact nonlocal star, i.e. $\epsilon\simeq 0.128,$ the compactness parameter is equal to $\mu_{\rm NL}\simeq 0.11,$ or in other words $r_{\rm sch}/r_{\rm NL}\simeq \sqrt{\pi}/2\simeq 0.886...\,.$ A very intriguing fact is that the compactness of a nonlocal star is of the same order of the Buchdahl limit \cite{Buchdahl:1959zz,Dadhich:2019jyf} which is $8/9\simeq 0.888...\,$ and was derived by Buchdahl by assuming a constant density interior. Indeed, in the nonlocal star when we take the limit $r\rightarrow 0$, the metric approaches conformal flatness with constant metric potentials or, equivalently, the effective Gaussian density source becomes constant in good approximation close to the origin.

\begin{itemize}

\item{\bf Nonlocal star as an ultracompact object}:
For a sufficiently compact nonlocal star we can have 
\begin{equation}\label{ultra}
r_{\rm NL}=2Gm(1+\epsilon)<3Gm,~~~0.128\lesssim \epsilon \lesssim 0.5\,,
\end{equation}
 which means that it can possess a photosphere, and can be seen as a new kind of {\it ultracompact object}. In this case, the properties of a nonlocal star can be probed by studying the ringdown phase. Indeed, after the merging process of two sufficiently compact objects there is a fraction of waves which will be able to cross the photosphere, but there will also be a fraction of them which will interact with the photosphere and travel back towards the central object. Since there is no horizon the waves can interact with the surface and travel back. This periodic behavior of a small fraction of gravitational waves would produce {\it echoes} in the wave-form signal \cite{Cardoso:2016oxy,Cardoso:2017cqb,Cardoso:2017njb,Carballo-Rubio:2018jzw}. 
 
 \item{\bf Echoes after ringdown phase}:
 In Ref.\cite{Cardoso:2017cqb}, an interesting distinction was made in the class of ultracompact objects, by distinguishing the so called {\it ClePhOs} (Clean Photosphere) from the non-ClePhOs. The main difference between these two kind of objects is that for the former echoes are produced at later times, and can be more easily distinguished by the rest of the wave-form by future LIGO/VIRGO observatories. ClePhOs have a radius $$R=r_{\rm sch}(1+\epsilon),~~~\epsilon\lesssim 0.0165.$$  From Eqs.(\ref{inequality},\ref{epsilon}), we can now understand that the nonlocal star is a particular class of ultracompact object which does not belong to the class of ClePhOs, since in its most compact case we have $\epsilon=0.128>0.01665.$

\item{\bf Shadow of a nonlocal star}: Moreover, since the object is sufficiently compact for Eq.~(\ref{ultra}), it will be possible to test and constrain the nonlocal star scenario by experimental data coming from the presence of a shadow~\cite{Falcke:1999pj}. In particular, we will be able to put constraints on the compactness parameter introduced in Eq.\eqref{function-compact} or, in other words, on the size of the object; see Refs.\cite{Cardoso:2017cqb,Carballo-Rubio:2018jzw} for discussions on phenomenological aspects of astrophysical horizonless objects beyond GR. 

\item{\bf Nonlocal star as a dark giant star}: When 
$$\epsilon > 0.5,$$ the radius of the nonlocal star is such that it engulfs the photosphere, i.e.
$r_{\rm NL}>3Gm$, so that no light ring would be present. Such a gravitationally bound system would be very similar to a dark giant star and they will hardly radiate, which will become evident when discussing the life time of a nonlocal star and the number of quantum states such an object possesses, see below. In terms of compactness, they will be very similar to a neutron~\cite{Ozel:2016oaf}, or a boson star~\cite{Schunck:2003kk}, but with an equation of state $p=0$.

\end{itemize}


\subsection{Life time of a nonlocal star }

We now wish to compute the life time of a nonlocal star. As a first step we need to find an expression for the {\it escape energy} and {\it escape wave-length} of a graviton, so that we can understand under which conditions a graviton will escape from the condensate. By following the procedure in Ref.\cite{Dvali:2011aa}, for a nonlocal condensate of $N$ weakly interacting quanta, we can define the following collective interaction strength:
%
$\hbar N\alpha_g=\hbar N{L_p^2}/{L^2_{\rm eff}}$,
%
so that each graviton feels the following collective binding potential, which coincides with the escape energy,
\begin{equation}
E_{\rm esc}\sim \frac{\hbar N\alpha_g}{L_{\rm eff}}\sim \left(\frac{L_p}{L_s}\right)^2\frac{\hbar}{\sqrt{N}L_s}. \label{coll-potential}
\end{equation}
where we have taken $r\sim r_{\rm NL}.$
Therefore, a graviton can escape from the condensate if its energy exceeds the escape energy in Eq.\eqref{coll-potential}. We can also find the escape wave-length through the relation
\begin{equation}
E_{\rm esc}=\frac{\hbar}{\lambda_{\rm esc}}. \label{de-broglie-escape}
\end{equation}
Indeed, by equating Eq.\eqref{coll-potential} and Eq.\eqref{de-broglie-escape}, we obtain
\begin{equation}
\lambda_{\rm esc}\sim \left(\frac{L_s}{L_p}\right)^2 \sqrt{N}L_s. \label{wavelength-escape}
\end{equation}
It is clear that the easiest way for a graviton to escape is through a $2\rightarrow 2$ scattering process, in which the energy of one of the two gravitons exceeds the threshold given by the binding potential $E_{\rm esc}.$ We can also estimate the {\it escape rate}, $\Gamma$, for such a process, which will be approximatively given by the product of the collective coupling squared, $\alpha_g^2,$ of the characteristic energy scale, $E_{\rm esc},$ and of the combinatoric factor, $N(N-1),$ which in the limit $N\gg1$ can be well approximated by $N^2:$
\begin{equation}
\Gamma\sim \alpha_g^2 N^2E_{\rm esc}\sim \left(\frac{L_p}{L_s}\right)^6\frac{\hbar}{\sqrt{N}L_s}. \label{rate}
\end{equation}
While, the corresponding time-scale of the process is given by
%
\begin{equation}
\Delta t=\frac{\hbar}{\Gamma}\sim \left(\frac{L_s}{L_p}\right)^6\sqrt{N}L_s. \label{time-scale}
\end{equation}
We can now study how the mass of the nonlocal star changes in time, namely its time variation which is given by:
\begin{equation}
\frac{dm}{dt}=-\frac{E_{\rm esc}}{\Delta t}= -\frac{\Gamma}{\lambda_{\rm esc}}\sim -\left(\frac{L_p}{L_s}\right)^8\frac{\hbar}{NL_s^2}, \label{mass-variation}
\end{equation}
or by using Eq.\eqref{total mass.2}, we obtain the time variation of $N$:
\begin{equation}
\frac{dN}{dt}\sim -\left(\frac{L_p}{L_s}\right)^8\frac{1}{\sqrt{N}L_s}. \label{N-variation}
\end{equation}
By imposing $dN/dt\sim -N/\tau,$ we can obtain an estimation for the {\it life time} of the nonlocal star:
\begin{equation}
\tau\sim \left(\frac{L_s}{L_p}\right)^8 N^{3/2}L_s\sim \left(\frac{L_s}{L_p}\right)^8 \frac{L_s^4}{\hbar^3}m^3, \label{life-time}
\end{equation}
from which it is clear that, given $N$ gravitons, the life time of a nonlocal star is always larger than the life time of a Schwarzschild blackhole, $\tau_{\rm bh}=N^{3/2}L_p$ \cite{Dvali:2011aa}, indeed we have:
\begin{equation}
\tau=\left(\frac{L_s}{L_p}\right)^{9}\tau_{\rm bh}=\left(\frac{M_p}{M_s}\right)^{9}\tau_{\rm bh}. \label{tau/tau_{bh}}
\end{equation}
The Hawking evaporation time for a Schwarzschild blackhole was computed for the first time by Don Page in Ref.~\cite{Page:1976df} and is given by $ \tau_{\rm bh}=8.66\times 10^{-27}(m/{\rm gram})^3$s. It is clear that 
nonlocal stars are even more stable as compared to Schwarzschild blackholes. Indeed, in order to have a life time larger than the age of the Universe, $10^{17}$s, from Eq.\eqref{tau/tau_{bh}} we obtain the following bound on the mass:
\begin{equation}
m>10^{14}\left(\frac{M_s}{M_p}\right)^3{\rm g}. \label{mass-bound}
\end{equation}
For  $M_s\sim 10^{16}$GeV, nonlocal stars with masses $m > 10^{5}$g would live up to the age of the Universe $>10^{17}{\rm s},$ which explicitly shows that nonlocal stars can live longer than a Schwarzschild blackhole of same mass.

The above analysis also suggests that  nonlocal stars are very efficient for storing information. There is no Hawking information loss paradox, due to the absence of a horizon, but the information can be kept inside for a long time, until the nonlocal star starts losing a sufficient amount of its initial mass. The end stage of the evaporation will not cause any remnant like in the case of a Schwarzschild blackhole, due to the absence of any event horizon.  

Note that nonlocal stars do open up a new parameter space for a stable compact object. After primordial inflation (for a review see~\cite{Mazumdar:2010sa}), the inflationary perturbations can create primordial blackholes~\cite{Carr:1974nx} (for a recent review see~\cite{Carr:2018rid,Sasaki:2018dmp}), which can live long enough to act as a dark matter candidate; in particular, for a value of the mass larger than $10^{14}$g they can live longer than the age of the Universe; see Ref.~\cite{Carr:2016hva,Sasaki:2018dmp} for various astrophysical constraints for masses above $10^{14}$g~\cite{Carr:2016hva,Sasaki:2018dmp}. Very interestingly, the nonlocal stars ameliorate this strict bound on $10^{14}$g, in fact a smaller massive nonlocal star (smaller than $10^{14}$g) can live even longer than the age of the Universe depending of the fundamental scale of nonlocality. Future investigation will be needed along this direction.



\subsection{Gravitational Entropy}

One of the prime reasons why the evaporation rate of a nonlocal star is smaller than the one of a Schwarzschild blackhole is related to the number of available Bekenstein states~\cite{Bekenstein:1980jp}. The key role is played by the fact that the number of states which can be stored in a nonlocal condensate is always larger than the corresponding number in the case of a blackhole, as we will explain below. We will only make a rough estimation of the entropy and of the number of available states in the nonlocal star. 

Let us introduce a characteristic length scale $L$ such that $dx\sim L$ and $\partial_x\sim 1/L,$ so that the classical action in Eq.\eqref{quad-action} can be recast as
\begin{equation}
\begin{array}{rl}
S=& \displaystyle \frac{\hbar}{L_p^2}\int d^4x \sqrt{-g} \left\lbrace \mathcal{R}+\beta\mathcal{R}\mathcal{F}_1(\Box_s)\mathcal{R}+\cdots\right\rbrace  \\[3mm]
\sim & \displaystyle \hbar\left(\frac{L^2}{L_p^2}+\frac{L_s^2}{L_p^2}\right).
\end{array}\label{dimensional-action}
\end{equation}
The first piece is the Einstein-Hilbert term, while the quadratic curvature contribution is given by 
$(M_p/M_s)^2=(L_s/L_p)^2$. Note that the quadratic part of the action is scale invariant in the sense that it does not depend on the characteristic scale $L$.  

Let us consider a nonlocal star with mass $m,$ in the quantum framework introduced above. If we take the Schwarzschild radius as the characteristic scale, $L\sim 2Gm$, the form of the action in Eq.\eqref{dimensional-action} as a function of $N$ gravitons reads
\begin{equation}
S\sim \hbar\left(\frac{4G^2m^2}{L_p^2}+\frac{r_{\rm NL}^2}{L_p^2}\right)\equiv \hbar \, s,
\label{entropy-both-single}
\end{equation}
where $s$ is the entropy of the gravitational system. Note that, if $r_{\rm NL}<2Gm$ the contribution from the Einstein-Hilbert term dominates, and we recover the usual Bekenstein entropy of a Schwarzschild blackhole, $s_{\rm bh}\sim G^2m^2/L_p^2$.  When  $r_{\rm NL}>2Gm$ the quadratic part of the action dominates, and then we obtain:  
\begin{equation}
s\sim \frac{L_{\rm eff}^2}{L_p^2}=N\frac{L_s^2}{L_p^2}=N\frac{M_p^2}{M_s^2}.
\label{entropy-nonlocal star}
\end{equation}
Note that an {\it Area-law} for the gravitational entropy still holds, $s\sim {\it Area}/4G$~\cite{Conroy:2015wfa}, but now the area is given by $4\pi r^2_{\rm NL}$. As a consistency check, note that if we set $M_s=M_p$, we recover the Bekenstein entropy, $s_{\rm bh}\sim N\sim Gm^2,$ consistently with the quantum corpuscular picture of a Schwarzschild blackhole \cite{Dvali:2011aa}.

The entropy of a nonlocal star turns out to be larger than the entropy of a Schwarzschild blackhole, $s>s_{\rm bh}$. As a consequence the number of available states contained in a nonlocal star is larger than the one in the blackhole case, indeed we have
\begin{equation}
\mathcal{N}\sim e^{N(L_s/L_p)^2}=e^{N(M_p/M_s)^2},
\label{states-nonlocal star}
\end{equation}
which is always larger than $\mathcal{N}_{\rm bh}\sim e^N.$ 

Having a larger number of states means that the probability for a graviton to escape from the condensate is lower, as the phase space of all possible configurations which can be realized is larger than that of  the blackhole case. Therefore, we can now understand why a nonlocal star can live longer than a blackhole. Although nonlocality makes the gravitational interaction weaker and the collective quantum coupling always smaller than one, Eq.\eqref{self-coup-less-nonlocal}, a gravitational bound state can be formed and a nonlocal star can live for a sufficiently long time, due to the fact that the number of states which can be occupied by the $N$ gravitons is sufficiently large.  The same reason makes it also an almost perfect absorber. Suppose, a high energy particle is absorbed by the nonlocal star, then the energy of the particle gets distributed amongst all the Bekenstein states, very similar to any blackbody system. 


\subsection{Absorption coefficient}
\begin{table*}[!t]
	\caption{Blackhole vs the most compact nonlocal star ($\epsilon\simeq 0.128$).}
	\centering
	\begin{tabular}{p{0.13\linewidth}p{0.125\linewidth}p{0.125\linewidth}p{0.12\linewidth}p{0.09\linewidth}p{0.15\linewidth}p{0.125\linewidth}}
		\toprule[1pt]\midrule[0.3pt]
		& radius & horizon &photosphere & $\mu$ & absorption  &life time  \\
		\midrule[0.3pt]
		blackhole & $2Gm$ & YES  & YES & 0 & 1 & $\displaystyle L_p^4\frac{m^3}{\hbar^3}$  \\
		nonlocal star & $2Gm(1+\epsilon)$ & NO & YES & 0.11 & $0.977\lesssim \kappa \lesssim 1$ &$\displaystyle \left(\frac{L_s}{L_p}\right)^8L_s^4\frac{m^3}{\hbar^3}$ \\
		\midrule[0.3pt]\bottomrule[1pt]
	\end{tabular}
\end{table*}
We now wish to obtain an estimation of the absorption coefficient of a nonlocal star in the most compact case, $\epsilon \simeq 0.128.$ We can do it by studying the dynamics of the system composed by the compact object and the accretion disk, see~\cite{Abramowicz:2011xu}, and using the experimental bounds on the exchanged fluxes of energies $\dot{M}_{\rm disk}$ and $\dot{E}.$ The former describes the amount of infalling matter coming from the accretion disk and going inside the central object per unit time, while the latter corresponds to the emitted energy from the compact object per unit of time. We can immediately notice that a classical blackhole is characterized by $\dot{E}_{\rm bh}=0$, as nothing can escape outside beyond the horizon.

First, note that one can compute which is the solid angle $\Delta \Omega$ under which particles coming out of the compact object can escape at infinity, and it is proportional to the parameter $\epsilon$, see \cite{Cardoso:2017njb}:
\begin{equation}
\frac{\Delta\Omega}{2\pi}=\frac{27}{8}\epsilon+\mathcal{O}(\epsilon^2).
\label{solid-angle}
\end{equation}
The last equation tells us that the more compact the object is, the more deflected the geodesics are, i.e. the more black the central object appears. This result is crucial in order to study the scenario in which a compact object is surrounded by an accretion disk.

Let us now introduce the following fundamental quantities for a nonlocal star: 
\begin{itemize}
\item $\kappa$ is the {\it absorption coefficient} and measures the amount of energy that is lost inside the nonlocal star.

\item $\gamma$ is the {\it elastic reflection coefficient} which measures the fraction of energy that reaches and interact elastically with the surface of the nonlocal star, and then is reflected back.

\item $\tilde{\gamma}$ is the {\it inelastic reflection coefficient} which measures the fraction of energy which is reflected after inelastic interaction with the surface, i.e. the portion of energy that goes inside the nonlocal star and subsequently is re-emitted.
\end{itemize}
 Note that the following relation holds: $\tilde{\gamma}=1-\kappa-\gamma.$ For a classical blackhole we have: $\kappa_{\rm bh}=1,$ $\gamma_{\rm bh}=\tilde{\gamma}_{\rm bh}=0.$ See also Ref.\cite{Carballo-Rubio:2018jzw}, for the definition of other fundamental parameters characterizing compact objects.

In the case of a compact object the flux of emitted energy can be non-vanishing, indeed the following relation holds true \cite{Carballo-Rubio:2018jzw}:
\begin{equation}
\frac{\dot{E}}{\dot{M}_{\rm disk}}\simeq \frac{(1-\kappa-\gamma)(1-\gamma)\Delta\Omega/2\pi}{\kappa+(1-\kappa-\gamma)\Delta\Omega/2\pi}.
\label{relation-absorp}
\end{equation}

Very interestingly, the quantity in Eq.\eqref{relation-absorp} can be constrained through astronomical observations; indeed, one can put the following upper bound \cite{Broderick:2007ek,Broderick:2007ek,Narayan:2008bv,Broderick:2009ph,Carballo-Rubio:2018jzw}:
\begin{equation}
\frac{(1-\kappa-\gamma)(1-\gamma)\Delta\Omega/2\pi}{\kappa+(1-\kappa-\gamma)\Delta\Omega/2\pi}\lesssim \mathcal{O}(10^{-2}).
\label{uppper-bound-rel}
\end{equation}
We are mainly interested in obtaining a direct bound on the absorption coefficient, but the way Eq.\eqref{uppper-bound-rel} is written would only allow us to constrain the combination of the two parameters $\kappa$ and $\gamma.$ However, we will still have to use one of the main features due to nonlocality: any particle falling into the nonlocal star gets lost for a long time, due to the fact that the number of states is exponentially large, see Eq.~(\ref{states-nonlocal star}). This means that a nonlocal star does {\it not} have a hard surface, and the corresponding elastic reflection coefficient is basically zero, $\gamma\approx 0.$ However, there is still a non-vanishing inelastic reflection coefficient which measures the fraction of infalling quanta which can come out again and escape from the nonlocal star, but we expect it to be very small by virtue of our previous results.

Therefore, for a nonlocal star Eq.\eqref{uppper-bound-rel} becomes
\begin{equation}
\frac{(1-\kappa)\Delta\Omega/2\pi}{\kappa+(1-\kappa)\Delta\Omega/2\pi}\lesssim \mathcal{O}(10^{-2}),
\label{uppper-bound-rel-2}
\end{equation}
from which, by using the formula Eq.\eqref{solid-angle} and working in the most compact case, $\epsilon\simeq 0.128,$ we get a direct bound on the absorption coefficient: 
\begin{equation}
0.977\lesssim \kappa \lesssim 1.
\label{bound-absorp}
\end{equation}
This result is consistent with the discussions on life time, entropy and number of states made above, indeed as expected the value of the absorption coefficient of a nonlocal star is very close to unity. As a consequence, since $\gamma\approx 0,$ we also obtain an upper bound on the inelastic reflection coefficient:
\begin{equation}
0\lesssim \tilde{\gamma} \lesssim 0.023,
\label{bound-inelast-coeff}
\end{equation}
which turns out to be very small, consistently with our expectations.


\section{Conclusions}\label{concl}

We have provided a quantum mechanical framework where we have described a non-singular, horizonless nonlocal star, for which the gravitational 
potential is always bounded below unity. The graviton interaction weakens due to the presence of nonlocal form-factors made up of infinite order covariant derivatives. For the simplest choice of an entire function, which appears in the graviton propagator, it is possible to avoid the curvature singularity and the event horizon, provided that the complementarity principle holds, $mM_{\rm eff}\sim M_s^2 < M_p^2$. In our case, the radius of the most compact nonlocal star is given by Eq.~\eqref{inequality}, 
i.e. $r_{\rm NL}=r_{\rm sch}(1+\epsilon),$ where $\epsilon\simeq 0.128,$ and such astrophysical objects can be excellent blackhole mimickers. Indeed, if $\epsilon < 0.5$, there is still a photosphere and echoes can be produced; therefore, their detection can allow to constraint the compactness parameter, which in turn would directly constrain the radius $r_{\rm NL}$. In this respect nonlocal stars can also be compared with fuzz-balls constructed via string states~\cite{Mathur:2005zp}, see Ref.~\cite{Guo:2017jmi} for a recent discussion. 

If the $\epsilon > 0.5$, then nonlocal stars could be compared with neutron and/or boson stars. However, the main difference is that now the equation of state parameter is characterized by a zero matter pressure, i.e. $p=0,$ but there could still be an effective pressure component coming from the gravity sector and related to the nonlocal geometry.

Nonlocal stars are very efficient absorbers and their life time is even longer than the one of a Schwarzschild blackhole, i.e.  $\tau/\tau_{\rm bh}\sim \left(L_s/L_p\right)^{9}.$ This is mainly due to the fact that the number of available states for a nonlocal star  is much larger than the corresponding number in the Schwarzschild blackhole case. It is this huge number of available states that compensates the weakness of the graviton interaction. A nonlocal star does not have a hard surface, so that no particles can be elastically backscattered. Moreover, by using experimental data related to the dynamics of the accretion disk surrounding a compact object, we have been able to put a lower bound on the absorption coefficient, $0.977\lesssim \kappa \lesssim 1.$  See Table I, for a summary and the comparisons between a classical blackhole and the most compact nonlocal star.

Finally, we wish to emphasize that we have only worked with the trace part of the cubic graviton action, see Eq.\eqref{cubic-graviton interac}. Nevertheless,  we believe  that the quantitative picture would still hold true even when the full tensorial structure, i.e. the interactions derived 
from the tensor part, is taken into account. Both the scalar and the tensor part of the graviton share the same exponentially suppressed propagator in the UV, therefore most of our results would still hold true for the spin-2 mode as well. Nevertheless, it would be desirable to study the interactions of both spin-2 and spin-0 components in full glory and study their quantum behavior to complete the picture of the nonlocal star. Further investigations will be subject of future works. 

Furthermore, from a physical point of view it will be very important to repeat a similar study in the case of rotating non-singular compact objects in infinite derivative gravity. Such kind of metric solutions have been studied~\cite{Buoninfante:2018xif}, but the quantum properties and the compactness are yet to be discussed in the literature.

\acknowledgements The authors are grateful to Sougato Bose, Francesco Di Filippo, Anish Ghoshal, Tanja Hinderer, Sravan Kumar, Gaetano Lambiase,  Stefano Liberati, Nobuchika Okada, Paolo Pani and Misao Sasaki for helpful discussions.  AM's research is financially supported by Netherlands Organization for Scientific Research (NWO) grant number 680-91-119.

\appendix

\section{Shifting the scale of nonlocality}\label{app-transmut}

In this Appendix we want to review how the fundamental scale of nonlocality $M_s$ can shift towards the IR regime as a function of the number of interacting particles $N$ \cite{Buoninfante:2018gce}. Let us work in Euclidean space and consider the following scalar action which can mimic the trace part of the graviton action up to order $\mathcal{O}(\kappa h^3)$~\cite{Talaganis:2014ida}:
\begin{equation}
\begin{array}{rl}
S=& \displaystyle \int d^4x \left( \frac{1}{2} h(x)e^{\left(-\Box/M_s^2\right)^n}\Box h(x)\right.\\[2.8mm]
& +\displaystyle\frac{\lambda}{4}h(x)\partial_{\mu}h(x)\partial^{\mu}h(x)\\[2.8mm]
&\displaystyle+\frac{\lambda}{4}h(x)\Box h(x)e^{\left(-\Box/M_s^2\right)^n}h(x)  \\[2.8mm]
&\displaystyle\left.-\frac{\lambda}{4}h(x)\partial_{\mu}h(x)e^{\left(-\Box/M_s^2\right)^n}\partial^{\mu}h(x)\right),
\end{array}\label{action-toy}
\end{equation}
where $\lambda$ is a coupling constant with inverse energy dimension. The propagator for this action is given by
\begin{equation}
\Pi(k)=\frac{e^{-(k^2/M_s^2)^n}}{k^2},
\end{equation}
while the interaction vertex reads \cite{Talaganis:2014ida,Talaganis:2016ovm}
\begin{equation}
\!\!\begin{array}{rl}
V(k_1,k_2,k_3)=&\displaystyle\frac{\lambda}{4}\left(k_1^2+k_2^2+k_3^2\right)\left(e^{k^{2n}_1/M_s^{2n}}+e^{k^{2n}_2/M_s^{2n}}\right.\\
&\displaystyle\,\,\,\,\,\,\,\,\,\,\,\,\,\,\,\,\,\,\,\,\,\,\,\,\,\,\,\,\,\,\,\,\,\,\,\,\,\,\,\,\,\,\,\,\,\,\,\left.+e^{k^{2n}_3/M_s^{2n}}-1\right).
\end{array}
\label{vertex-toy}
\end{equation}
One can show that both propagator and vertex become exponentially suppressed once they are dressed by taking into account of quantum corrections \cite{Talaganis:2014ida,Talaganis:2016ovm,Buoninfante:2018mre,Buoninfante:2018gce}.

We now want to compute an $N$-point scattering amplitude with, which is in one-to-one correspondence with the physics of a Bose-Einstein condensate; in particular, as an example, we can consider a tree-level amplitude, made up of $N$ external legs, $N-2$ vertices and $N-3$ internal propagators, which we assume to be dressed. By taking the limit $N\gg 1$ one can notice that such an $N$-point amplitude has the following behavior \cite{Buoninfante:2018gce}:
\begin{equation}
\mathcal{M}_N\sim \frac{e^{-N^{2n+1}k^2/M_s^2}}{k^{2N}}= \frac{e^{-K^2/M_{\rm eff}^2}}{k^{2N}},\label{ampl-1-loop-deriv}
\end{equation}
where we have neglected constant factors and used the relations 
\begin{equation}
p_1+p_2+\cdots +p_{N-2}=p_{N-1}+p_{N},\label{cons-law-N}
\end{equation}
\begin{equation}
|p_i+p_{i+1}|\equiv|p|,\,\,\,\,\vec{p}_i=-\vec{p}_{i+1},\,\,\,\, i=1,\dots,N-3;
\label{N-point-choice}
\end{equation}
and 
\begin{equation}
2j^2p^2\gg (p_{2j+1}^4)^2,\,2p^4_{2j+1}jp\,.\label{neglect-condition}
\end{equation}
From Eq.\eqref{ampl-1-loop-deriv}, by defining the total momentum (energy) thrown into the system as $K=Nk,$ it is clear that the fundamental scale of nonlocality {\it effectively} transmutes to lower energies:
\begin{equation}
M_{\rm eff}= \frac{M_s}{N^{\frac{1}{2n}}}.\label{eff-scale-zero-n-deriv}
\end{equation}
In this paper, we have constructed a model of nonlocal star for $n=1,$ which corresponds to $M_{\rm eff}= M_s/\sqrt{N}.$ We wish to mention that for any other positive integer $n>1,$ the same procedure does not work as the horizon may be still formed. 

In fact, if we had considered the general entire function 
\begin{equation}
e^{\gamma(\Box_s)} =  e^{\left(-\Box_s\right)^n},
\end{equation}
with $n>0$, we would have obtained $M_{\rm eff}={M_s}/{N^{1/2n}}$, or $L_{\rm eff}=N^{1/2n}L_s$. Furthermore, the total mass of an object with $N$ gravitons would scale as $M\sim N^{(2n-1)/(2n)}M_s$, and the complementarity relation 
between mass and scale of nonlocality would have been modified by $MM_{\rm eff}\sim N^{({n-1)}/(n)}M_s^2$. As a consequence, the collective quantum coupling of $N$ scalar gravitons would read:
$N\alpha_g= N{L_p^2}/{L_{\rm eff}^2}=N^{(n-1)/n}(L_p/L_{s})^2$, which means that, for a sufficiently large $N,$ it  can assume a value equal to or larger than one, for any $n>1$.
Therefore, the only power of $\Box^{n}$ which does not violate the inequalities in Eqs.(\ref{inequality} (or equivalently Eq.\eqref{inequality-always}), and can always prevent the formation of an event horizon, is $n=1.$

\end{document}